\documentclass[12pt]{article}
\usepackage{amssymb}
\usepackage{epsfig,caption,graphicx,eepic,epic}
\usepackage{pstricks}
\usepackage{amsmath}
\usepackage{multido}

\begin{document}
\def\be{\begin{equation}}
\def\ee{\end{equation}}
\def\ba{\begin{eqnarray}} 
\def\ea{\end{eqnarray}}
\def\nn{\nonumber}

\newcommand{\bbf}{\mathbf}
\newcommand{\rrm}{\mathrm}

\title{Low-energy properties of non-perturbative quantum systems: a space reduction 
approach\\} 


\author{Tarek Khalil$^{a,b}$
\footnote{E-mail address: khalil@lpt1.u-strasbg.fr}\\ 
and\\
Jean Richert$^{a}$
\footnote{E-mail address: richert@lpt1.u-strasbg.fr}\\ 
$^{a}$ Laboratoire de Physique Th\'eorique, UMR 7085 CNRS/ULP,\\
Universit\'e Louis Pasteur, 67084 Strasbourg Cedex,\\ 
France\\
$^{b}$ D\'epartement de Physique, Facult\'e des Sciences, Section V,\\
Universit\'e Libanaise, Nabatieh,\\ 
Liban} 
 
\date{\today}
\maketitle 
\begin{abstract}
We propose and test a renormalization procedure which acts in Hilbert space. We test its 
efficiency on strongly correlated quantum spin systems by working out and analyzing the 
low-energy spectral properties of frustrated quantum spin systems in different parts of the 
phase diagram and in the neighbourhood of quantum critical points.
\end{abstract} 
\maketitle

{\it Introduction.}\\

Microscopic many-body quantum systems are often subject to strong interactions which act 
between their constituents. Perturbation treatments make sense if it is possible to introduce 
a mean-field which is able to absorb the main part of the interaction between the constituents.
But often such an approach does not lead to sensible results, in particular in the case of  
realistic quantum spin systems. Non-perturbative procedures are needed, see f.i. ~\cite{pol,vid,whi2}.
\vskip .1cm
Spectral properties of quantum systems are obtained by means of the diagonalization of a 
many-body Hamiltonian in Hilbert space which is spanned by a complete, in general infinite 
or at least a very large basis of states although the information of interest is restricted 
to the knowledge of a few low energy states. In order to reduce the problem to these states 
we proposed a new approach which acts as a size reduction of Hilbert space. 
The procedure consists of an algorithm which eliminates states by means of a step by step 
projection procedure to subspaces of the original space ~\cite{khri}. It relies on the general
renormalization concept ~\cite{wil,gla,mue,bek}. It can be applied to all types of microscopic 
quantum systems (molecular, atomic, nuclear, solid state,...) in contradistinction with more 
specific procedures such as f.i. the Density Matrix Renormalization Group ($DMRG$)
~\cite{whi2,jos,mal,car} which are more specifically applied to systems which live 
on a lattice. The  reduction concept shows some connection with the method which has been proposed and 
applied in problems related to the treatment of phonons~\cite{weib,fried} by means of an efficient 
method which connects a density matrix approach with optimized space dimension reduction.
\vskip .1cm
In the present work we implement this algorithm as a preliminary test of the practical efficiency 
and accuracy of the method when applied to large strongly interacting systems. We want to test  its 
capacity to deliver precise information about the low energy eigenstates of the many-body systems 
it is aimed to describe. Here we use quantum spin systems as test probes.

\vskip .2cm
{\it The reduction procedure.}
\vskip .2cm

We consider a system described by a Hamiltonian depending on a unique coupling strength 
$g$ which can be written as a sum of two terms 
\be
H = H_0 + g H_1
\label{eq0}   
\ee

The Hilbert space  ${\cal H}^{(N)}$ of dimension $N$ is spanned by a priori arbitrary complete 
set of basis states 
$\left\{|\Phi_i\rangle, \,i=1,\cdots, N\right\}$ which may be f.i. the eigenstates of $H_{0}$. 
Eigenvectors $|\Psi_j^{(N)}\rangle$
decompose on the basis as
\be
|\Psi_j^{(N)}\rangle = \sum_{i=1}^{N}  a_{ji}^{(N)}(g^{(N)})|\Phi_i\rangle
\label{eq1}   
\ee    
where the amplitudes $\{a_{ji}^{(N)}(g^{(N)})\}$ depend on the value $g^{(N)}$ of
$g$ in the chosen Hilbert space ${\cal H}^{(N)}$.
\vskip .1cm
The space can be decomposed into two subspaces by means of the projection operators $P$ 
and $Q$ ~\cite{fesh},
\be
{\cal H}^{(N)} = P{\cal H}^{(N)} + Q{\cal H}^{(N)}
\label{eq2}
\ee
The projected eigenvector $P|\Psi_j^{(N)}\rangle$ obeys then the effective Schr\"odinger 
equation  
\be
H_{eff}(\lambda_j^{(N)})P |\Psi_j^{(N)}\rangle =  \lambda_j^{(N)}
P |\Psi_j^{(N)}\rangle
\label{eq3}  
\ee
where $H_{eff}(\lambda_j^{(N)})$ is a new Hamiltonian which operates in the subspace  
$P{\cal H}^{(N)}$. It depends on the eigenvalue $\lambda_j^{(N)}$ which is also the eigenenergy 
corresponding to $|\Psi_j^{(N)}\rangle$ in the initial space ${\cal H}^{(N)}$. The coupling 
$g^{(N)}$ which characterizes the Hamiltonian $H^{(N)}$ in ${\cal H}^{(N)}$ is now constrained 
to change into $g^{(N-1)}$ in such a way that the eigenvalue in the new space ${\cal H}^{(N-1)}$ 
is the same as the one in the original space
\be
\lambda_j^{(N-1)} = \lambda_j^{(N)} 
\label{eq4} \  
\ee
The determination of $g^{(N-1)}$ by means of the constraint expressed by Eq.~(\ref{eq4}) is 
the central point of the procedure. It corresponds to a renormalization procedure which is
implemented by means a non-linear relation between $g^{(N)}$ and $g^{(N-1)}$ ~\cite{khri,kh}.
\vskip .1cm
In the sequel $P|\Psi_1^{(N)}\rangle$ is chosen to be projection of the ground state 
eigenvector $|\Psi_1^{(N)}\rangle$ and $\lambda_1^{(N)} = \lambda_1^{(N-1)} = \lambda_1$ the 
corresponding ground state energy. Following this 
first operation the reduction procedure is iterated in a step by step decrease of the dimensions 
of Hilbert space, $N \mapsto N-1 \mapsto N-2 \mapsto...$ leading at each step $k$ to a new 
coupling strength $g^{(N-k)}$ which is fixed as the solution of an algebraic equation and is shown 
to obey a flow equation in the limit of a continuum description ~\cite{khri,kh}. 

\vskip .2cm
{\it The implementation of the reduction algorithm}
\vskip .2cm

The procedure goes along the following steps:

$1-$ Consider a quantum system described by an Hamiltonian $H^{(N)}$ and compute its matrix elements 
in a definite basis of states $\{|\Phi_i\rangle, i=1,\ldots,N \}$. 
\vskip .1cm
$2-$ Use the Lanczos technique to determine $\lambda_1^{(N)}$ and the amplitudes $a_{1i}^{(N)}(g^{(N)})$ 
corresponding to $|\Psi_1^{(N)}(g^{(N)})\rangle$~\cite{lanc1}. The diagonal matrix elements 
$\{\epsilon_i = \langle \Phi_i|H^{N}| \Phi_i \rangle\}$ are arranged in decreasing order of values 
of the $|a_{1i}^{(N)}(g^{(N)})|$.
\vskip .1cm
$3-$ Fix $g^{(N-1)}$ as described above and in ~\cite{khri}. Construct 
$H^{(N-1)} = H_0 + g^{(N-1)} H_1$ by elimination of the matrix elements of 
$H^{(N)}$ involving the state $|\Phi_N\rangle$.
\vskip .1cm
$4-$ Repeat the procedures $2$ and $3$ by fixing at each step $k$
$\lambda_1^{(N-k)}=\lambda_1^{(N)} = \lambda_1$. The iterations are stopped at $N = N_{min}$ which may 
correspond to the limit of space dimensions for which the spectrum gets unstable.

\vskip .2cm 
This is due to the fact that $|\Psi_1^{(N-k-1)}\rangle$ which is the 
eigenvector in the space ${\cal H}^{(N-k-1)}$ 
and the projected state $P|\Psi_1^{(N-k)}\rangle$ of $|\Psi_1^{(N-k)}\rangle$ into 
${\cal H}^{(N-k-1)}$ may not coincide exactly. As a consequence it may not be possible to keep  
$\lambda_1^{(N-k-1)}$ rigorously equal to $\lambda_1^{(N-k)} = \lambda_1$. In practice the degree 
of accuracy depends on the relative size of the eliminated amplitudes 
$\{a_{1(N-k)}^{(N-k)}(g^{(N-k)})\}$.
This point will be tested by means of numerical estimations and further discussed below.

\vskip .2cm
{\it Quantum Phase transitions and fixed points}
\vskip .2cm 
 
Strongly correlated systems often possess rich phase diagrams and critical transition points
~\cite{sach1,aso,mvoj}. We 
show now how our procedure reflects the presence of these transitions.
\vskip .1cm
The eigenvalues $\lambda_k{(g)}$ of $H(g) = H_0 + gH_1$ are analytic functions of $g$ which may 
show algebraic singularities~\cite{kat,hei,sch2} at so called exceptional points $g = g_e$. Exceptional
points are first order branch points in the complex $g$ - plane which appear when two (or more) 
eigenvalues get degenerate. This can happen if $g$ takes values such that $\epsilon_k =  \epsilon_l$ 
where $\epsilon_k  = \langle \Phi_k|H|\Phi_k\rangle$. In a finite Hilbert space the degeneracy appears
as an avoided crossing for real $g$. If an energy level $\epsilon_k$ belonging to the $P{\cal H}$ 
subspace defined above crosses an energy level $\epsilon_l $ lying in the complementary $Q{\cal H}$ 
subspace the perturbation development constructed from $H_{eff}(E)$ diverges~\cite{sch2}. 
Physical states can get degenerate in energy for real values of $g_e$.  

Exceptional points are defined as the solutions of
\be 
f(\lambda(g_e))  = det[ H(g_e) - \lambda(g_e)I] = 0
\label{eq5} 
\ee
and 
\be
\frac{df(\lambda(g_e))}{d\lambda}|_{\lambda= \lambda(g_e)}= 0
\label{eq6} 
\ee

They are fixed points of the coupling strength $g$ which stays constant during the 
space reduction process ~\cite{khri,kh}. Indeed, if $\left\{\lambda_i(g)\right\}$ are the set of 
eigenvalues the secular equation can be written as
\be
\prod_{i=1}^N {(\lambda - \lambda_i)} = 0
\label{eq7} 
\ee
Consider $\lambda = \lambda_p$ which satisfies Eq.~(\ref{eq5}). Eq.~(\ref{eq6}) can only be 
satisfied if there exists another eigenvalue $\lambda_q = \lambda_ p$, hence if a degeneracy 
appears in the spectrum. This is the case at an exceptional point~\cite{kat}.
\vskip .1cm 
If the eigenvalue $\lambda_j^{(N-k)}, k= 0, 1,...$ which is either constant or constrained to take 
a fixed value $\lambda_j$ gets degenerate with some other eigenvalue 
$\lambda_i^{(N-k)}({g = g_e})$  in the space reduction process this eigenvalue must satisfy 
\be
\lambda_i^{(N-k)}({g_e}) = \lambda_i^{(N-l)}({g'_e})
\label{eq8}  
\ee
at any step $k$ and $l$ of the projection procedure. In the continuum limit for 
large values of $N$, $(N, N-1) \rightarrow (x, x-dx)$, 

\be
\frac{d\lambda_j}{dx} = 0 = \frac{d\lambda_i(g_e(x),x)}{dx}
\label{eq9} \ 
\ee 
 
where 
 
\be\notag \\
\lambda_i(g_e(x),x) = \langle\Psi_{i}(g_e(x),x)|H(g_e(x))|\Psi_{i}(g_e(x),x)\rangle 
\ee

Consequently
\be
\frac{\partial \lambda_i}{\partial x} + \frac{\partial \lambda_i}{\partial g_e} 
\frac{dg_e}{dx} = 0
\label{eq10} \ 
\ee

Since $\lambda_i(g_e(x),x)$ stays constant with $x$ in the space dimension interval $(x-dx,x)$ the 
first term in Eq.~(\ref{eq10}) is equal to zero. Hence in general

\be
\frac{\partial \lambda_i}{\partial g_e}\not = 0 \,\,\, \quad {\mathrm{and}}\,\,\,\,\,
\frac{dg_e}{dx} = 0
\label{eq11}  
\ee  

which shows that the exceptional point $g = g_{e}$ corresponds to a fixed point in the renormalization 
process, and characterizes the existence of a quantum phase transition.  
\vskip .1cm
This is general and verified for any Hamiltonian for which state degeneracies occurs. Numerical 
applications corresponding to a linear dependence on $g$, $H(g) = H_0 + g H_1$ follow.

\vskip .2cm
{\it Applications to a frustrated two-leg quantum spin ladder.}
\vskip .2cm 
Quantum spin systems are strongly interacting systems whose properties cannot be studied by means of
perturbation expansions. Hence they are particularly well adapted to non-perturbative treatments such as
renormalization procedures. In the following we present a test of the present method on such a system.

\begin{figure}
\epsfig{file=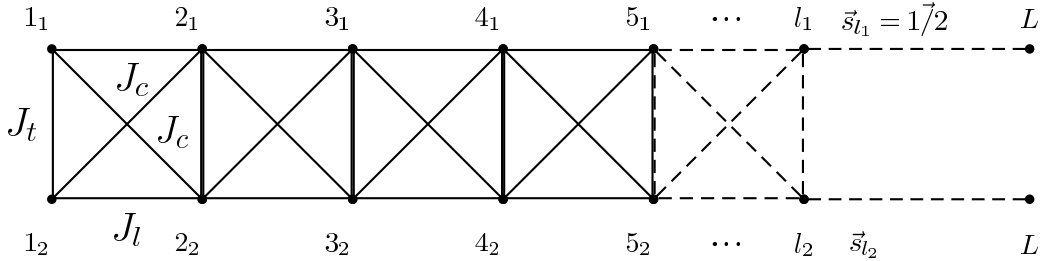}
\caption{The frustrated spin ladder. The coupling strengths are indicated as given in the text.}
\label{fig1}
\end{figure}

Consider a spin-$1/2$ ladder as shown in Figure \ref{fig1}. Its Hamiltonian $H(g) = H_0 + gH_1$
corresponds to $H_0 = 0$, $g=g^{(N)} = J_t$ and 
\be\label{eq12} 
H_1 = \sum_{i=1}^{L} s_{i_1}s_{i_2} + \gamma_{tl} \sum_{<ij>}(s_{i_1}s_{j_1} +
s_{i_2}s_{j_2})  + \gamma_{c}\sum_{<ij>}(s_{i_1}s_{j_2} + s_{i_2}s_{j_1}) 
\ee 
where $\gamma_{tl} = J_{l}/J_{t}$, $\gamma_{c} = J_{c}/J_{t}$ which are kept constant and $g^{(N)}$
is the renormalizable coupling strength. The number of sites along a leg is $L$. Indices $1$ or $2$ 
label the vector spin $1/2$ operators $s_{i_k}$ acting on the sites $i$ of leg $k$, $<ij>$ labels 
nearest neighbours on a leg, $(ij)$ diagonal interactions between sites located on different 
legs. The coupling strengths $J_t, J_l, J_{c}$ are positive. One can show that the renormalization 
process may be indifferently realized on $J_l$ or $J_c$ and 
leads to the same result~\cite{kh}.

\vskip .1cm
The basis states $\{|\Phi_k\rangle$, $k=1,\ldots,N\}$ are generated as product states 
 $|1/2 ~~m_i\rangle$

\be\nonumber
|\Phi_k\rangle = |1/2 ~~m_1;...,1/2 ~~m_i;...,1/2 ~~m_{2L}, \sum_{i=1}^{2L} m_i= M_{tot}\rangle  
\ee
with $\{m_i = +1/2, -1/2\}$. The total projection is a good quantum number fixed to $M_{tot} = 0$
in the following.

\vskip .2cm
{\it Test observables.}
\vskip .2cm 

In order to estimate the accuracy of the reduction procedure we introduce the quantity

\ba\label{eq13}
p(i) = |\frac{(e_i^{(N)}-e_i^{(n)})}{e_i^{(N)}}| \times 100 &with& i=1,\ldots,3
\ea 
where $e_i^{(n)} = \lambda_i^{(n)}/2L$ with  $n=(N-k)$ corresponds to the energy per 
site at the $i$th physical state starting from the ground state at the $k$th iteration in Hilbert 
space. This quantity provides a percentage of loss of accuracy of the eigenenergies in the 
different reduced spaces.
\vskip .1cm
A global characterization of the ground state wavefunction can also 
be given by the entropy per site in a space of dimension $n$

\ba\label{14}
s^{(n)} = - \frac{1}{2L} \sum_{i=1}^{n}{P_i}ln{P_i} & with & P_i = |\langle\Phi_i|\Psi_1^{(n)}
\rangle|^{2} = |a_{1i}^{(n)}|^{2}
\ea
which works as a measure of the distribution of the amplitudes $\{a_{1i}^{(n)}\}$ in the physical 
ground state~\cite{sok}.

\vskip .2cm
{\it Results and discussion.}
\vskip .2cm

We apply the reduction algorithm to the two-leg ladders introduced above for different numbers of 
sites and different values of the coupling strengths $J_{t}$, $J_{l}$ and $J_{c}$ in the $M_{tot}=0$ 
subspace.

\vskip .2cm
{\it First test: $L$= 9, $J_t$=15 and 2.5, $J_l$=5, $J_{c}$=3}
\vskip .2cm 

\begin{figure}[ht]
\includegraphics{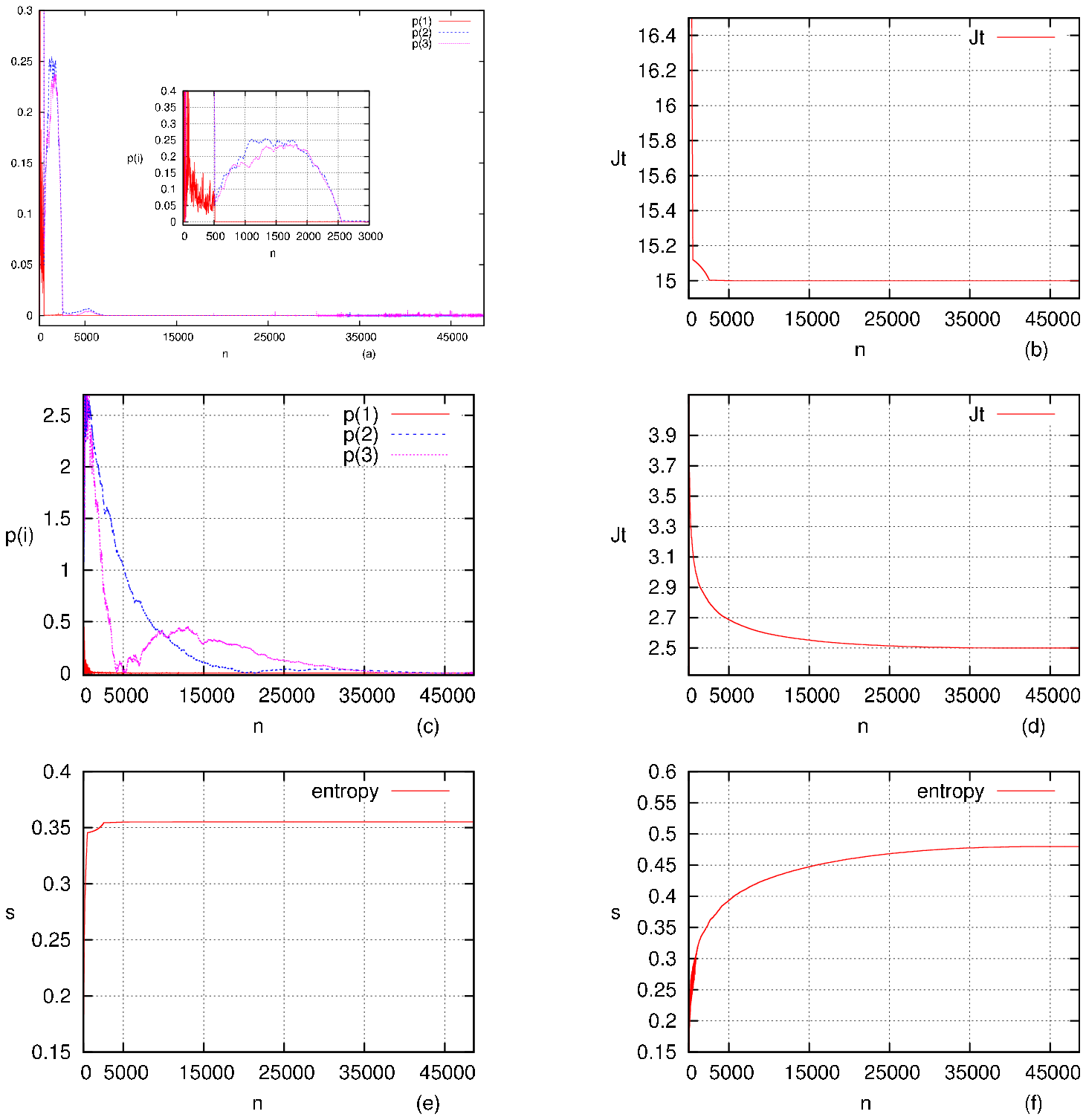}
\caption{Behaviour of $\{p_i, i= 1,2,3\}$, $J_{t}$ and the entropy $s$ as a function of the size 
$n$ of the Hilbert space. Cases $(a)-(b)$ and $(e)$ correspond to $J_{t}=15$, $(c)-(d)$ and $(f)$ 
to $J_{t}=2.5$, $J_l=5$, $J_{c}=3$. The number of sites along a leg is $L=9$.}
\label{fig2}
\end{figure}

Results for $L=9$ sites along a leg in the Hilbert space spanned by $N=48620$ basis states with 
$M_{tot}=0$ are presented in Figs.(\ref{fig2}(a-f)):

\begin{itemize}

\item For $J_t=15$, Fig.(\ref{fig2}(a)) shows that $p(1)$ increases from very small values to 
$0.5 \%$ when $n$ lies between $500$ and $80$, $p(2)$ and $p(3)\simeq 0.05 \%$ when $n \simeq 500$.\\

\item For $J_t=2.5$, Fig. (\ref{fig2}(c)) shows that $p(1)$ increases also from very small values 
to $0.5 \%$ when $n$ lies between $1000$ and $100$, $p(2)$  and $p(3)\simeq 2.5 \%$ when 
$n \simeq 500$.\\

\end{itemize}

Instabilities in the position of the eigenstates are localized at dimensions of the reduced Hilbert 
space where the elimination of components of the wave function cannot be properly compensated by 
the renormalization of $g = J_{t}$ which changes sizably at these values. This can be shown by the close 
relation in reduced Hilbert space between the behaviour of the renormalized coupling parameter 
$J_{t}^{(n)}$ and the entropy $s^{(n)}$ which works as a measure of the distribution of the amplitudes 
in the ground state, see Figs.(\ref{fig2}(b)-\ref{fig2}(e)) and (\ref{fig2}(d)-\ref{fig2}(f)). The 
effect of instability is weaker for $J_{t}=15$ which corresponds to a strong coupling along the rungs 
whereas $J_{t}=2.5$ corresponds to a stronger coupling along the legs of the ladder.
This can be explained by means of symmetry arguments~\cite{kh1,kh2}.
\vskip .2cm
{\it Second test: $L$= 6, $J_{t}$=15, $J_{l}\simeq 12.21$, $J_{c}\simeq 12.21$, $11$}
\vskip .2cm

\begin{figure}[ht]
\includegraphics{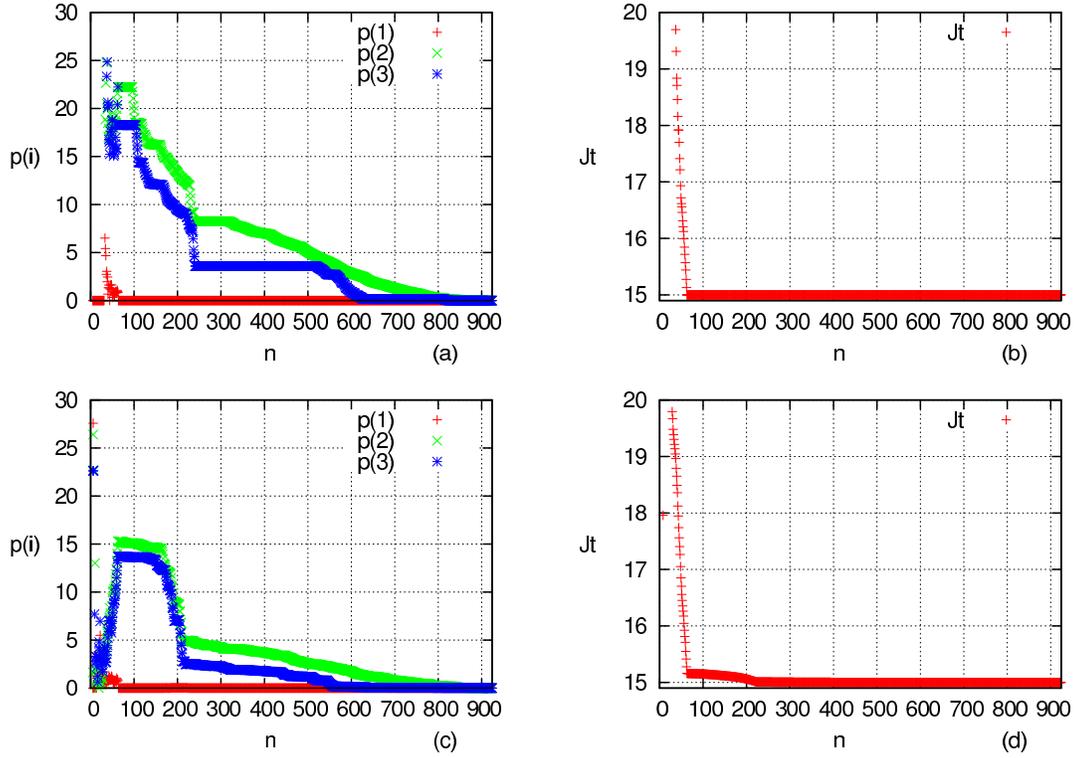}
\caption{Same notation as in Fig(\ref{fig2}). Cases $(a)-(b)$ correspond to 
$(J_{t}/J_{l})_{crit}\simeq 1.23$, $J_{t}=15,J_{l}=J_{c}\simeq 12.21$. Cases $(c)-(d)$ 
correspond to $J_{t}=15,J_{l}\simeq 12.21, J_{c}=11$. The number of sites along a leg is $L=6$. 
Broadened lines are drawn in order to facilitate the reading.}
\label{fig3}
\end{figure}

Here we test the behaviour of the system at a first order phase transition point.
\vskip .2cm
For $L=6$ sites along a leg the Hilbert space is spanned by $N=924$ basis states with $M_{tot}=0$. 
Results are shown in Figs(\ref{fig3}(a-d)).
\vskip .2cm
This case corresponds to the transition from a rung dimer phase to a Haldane phase which appears for 
$J_l = J_{c}$ when $(J_t/J_l)_{crit} \simeq 1.401$ in the case of an asymptotically large system
~\cite{gel}. The ratio depends on the size of the system. As predicted by the theory above the coupling 
constant $g = J_{t}$ is expected to stay constant at the level crossing point.
\vskip .2cm
For $J_{t}$=15, $J_{l}=J_{c}\simeq 12.21$, $e_{1}=-11.25,e_{2}=-11.25, e_{3}\simeq -10.7$. One observes 
a level crossing between the ground state and first excited state of the energy spectrum. 
Fig.(\ref{fig3}(b)) shows the constancy of the coupling strength down to low dimensions of $N<100$. 
For smaller $N$, when components of the wave function with sizable weight are eliminated,
$J_{t}$ increases sizably. In Fig.(\ref{fig3}(a)) one notices that the ground state remains stable 
but the first excited states move abruptly during the reduction procedure.
\vskip .2cm
For $J_{t}=15$, $J_{l}\simeq 12.21, J_{c}\simeq 11$, $e_{1}\simeq -11.38,e_{2}\simeq -10.67,
e_{3}\simeq -10.4$ one stays in the vicinity of the transition point. Figs.(\ref{fig3}(c-d)) show 
the behaviour of the spectrum and the coupling strength $J_{t}$. One notices that in  
Fig.(\ref{fig3}(d)) the renormalized coupling parameter stays stable down to $N<250$. In 
Fig.(\ref{fig3}(c)) the ground state remains stable but the first excited states goes moving during 
the reduction procedure but less than in the case shown in Fig.(\ref{fig3}(a)).

\vskip .2cm
{\it Summary - conclusions.}
\vskip .2cm
 
In the present work we tested and analysed the outcome of an algorithm which aims to reduce the 
dimensions of the Hilbert space of states describing strongly interacting systems. The reduction  
induces the renormalization of the coupling strengths which enter the Hamiltonians. By construction 
the algorithm works in any space dimension and may be applied to the study 
of any microscopic $N$-body quantum system.
\vskip .1cm
The outcome of numerical tests of the method applied to strongly correlated and frustrated quantum 
spin ladders can be summarized as follows:

\begin{itemize} 

\item Local spectral instabilities of the ground and low excited states appearing in the course of the 
reduction procedure developed above are correlated 
with the elimination of basis states with sizable amplitudes in the ground state wavefunction. The 
renormalization is able to cure this instability down to a small values of the Hilbert space dimensions. 
Numerical examples show that the procedure allows for sizable reduction of the dimensions of Hilbert 
space.

\item The stability of the low-lying states of the spectrum in the course of the reduction procedure 
depends on the relative strengths of the  coupling constants. Here the ladder favours a dimer structure 
(i.e. strong coupling along the rungs) for which the stability is the better the larger this coupling. 
Symmetry arguments can explain such a behaviour.

\item As it could be expected the system is unstable at phase transition points. The theory predicts 
a constant coupling parameter there. This is the case in numerical applications, down to a limit related to 
the space dimension reduction discussed above. The ground state stays remarkably stable, the spectrum 
of excited states gets however strongly unstable there.

\end{itemize} 

Last, the present procedure may be extended to the renormalization of more than one coupling 
parameter and possibly to systems which are not linear in $g$. However in most cases realistic 
microscopic systems and in particular quantum spin systems are described by Hamiltonians which depend
linearly on coupling strengths. The study of systems at finite temperature can also be performed
~\cite{jr}.

\end{document}